\documentclass[12pt]{article}

\usepackage{amsmath}
\usepackage{amssymb}

\def\rdots{\mathinner{\mkern1mu\raise1pt\vbox{\kern1pt\hbox{.}}\mkern2mu
   \raise4pt\hbox{.}\mkern2mu\raise7pt\hbox{.}\mkern1mu}}
\newcommand{\Z}{{\rm Z\kern-.35em Z}}
\newcommand{\bP}{{\rm I\kern-.15em P}}
\newcommand{\Q}{\kern.3em\rule{.07em}{.65em}\kern-.3em{\rm Q}}
\newcommand{\R}{{\rm I\kern-.15em R}}
\newcommand{\h}{{\rm I\kern-.15em H}}
\newcommand{\C}{\kern.3em\rule{.07em}{.65em}\kern-.3em{\rm C}}
\newcommand{\T}{{\rm T\kern-.35em T}}

\newcommand{\be}{\begin{equation}}
\newcommand{\ee}{\end{equation}}

\newcommand{\la}{\lambda}

\newcommand{\al}{\alpha}

\begin{document}

\openup 1.5\jot
\centerline{Dimer $\la_d$ Expansion Computer Computations}

\bigskip

\bigskip

\vspace{1in}
\centerline{Paul Federbush}
\centerline{Department of Mathematics}
\centerline{University of Michigan}
\centerline{Ann Arbor, MI 48109-1043}
\centerline{(pfed@umich.edu)}

\vspace{1in}

\centerline{\underline{Abstract}}

\ \ \ \ \ In a previous paper an asymptotic expansion for $\la_d$ in powers of $1/d$ was developed.  The results of computer computations for some terms in the expansion, as well as various quantities associated to the expansion, are herein presented.  The computations (in integer arithmetic) are actually done only for $d=1, d=2$, and $d=3$, but some results for arbitrary dimension follow from the general structure of the theory.  In particular we obtain the next term in powers of $1/d$, the $1/d^3$ term in the asymptotic expansion for $\la_d$, as well as checking the correctness of the terms previously calculated by hand.

For $d=2$ and $d=3$ a second expansion for $\la_d$ is defined and numerically studied.  Contrary to our hopes this second type of expansion appears not to be convergent.  There is not enough evidence to argue if the expansions for $\la_d$ in powers of $1/d$ are convergent for either $d=2$ or $d=3$, but likely they are not. ( Certainly convergence, but at a very slow rate, remains an interesting
possibility in all our cases. )

\vfill\eject 

\ \ \ \ \ In [1] an expansion for $\la_d$ in powers of $1/d$ was presented, completing the asymptotic expression obtained in [2], by Minc.  The expression as presented in [1] is
\be	\la_d \sim \frac 1 2 \; ln(2d) - \frac 1 2 + \frac 1 {8} \frac 1 d + \frac5{96} \; \frac1{d^2}  + \ \cdots  \ee
One result of this paper is the extension for one more power of $1/d$
\be	\la_d \sim \frac 1 2 \; ln(2d) - \frac 1 2 + \frac 1 {8} \frac 1 d + \frac5{96} \; \frac1{d^2}  + \frac 5{64}\;  \frac1{d^3} + \cdots. \ee
The key quantities to be computed are the $\bar J_i$ defined in equation (28) of [1], and the note after (33) therein
\be	\bar J_i = \frac 1 N \ \frac 1 {i!} \ J_i	\ee
The first few $\bar J_i$ are given in (29) and (38) of [1].

We have used Maple software to compute the $\bar J_i$ for $i=1,2,3,4,5,6; \ d=1,2,3$ as given in the following.

\noindent
\underline{For $d=1$ one has}:
\begin{eqnarray}
\bar J_1 &=& 0 \\
\bar J_2 &=&  1/8 \\
\bar J_3 &=& 1/12 \\
\bar J_4 &=& -\; 3/64 \\
\bar J_5 &=&  -\; 13/80\\
\bar J_6 &=&  -\; 19/192
\end{eqnarray} 
\underline{For $d=2$ one has}:
\begin{eqnarray}
\bar J_1 &=& 0 \\
\bar J_2 &=&  1/16 \\
\bar J_3 &=& 1/48 \\
 \bar J_4 &=& -\; 9/512 \\
\bar J_5 &=&  -\; 23/1280\\
\bar J_6 &=&  25/3072 
\end{eqnarray}
\underline{For $d=3$ one has}:
\begin{eqnarray}
\bar J_1 &=& 0 \\
\bar J_2 &=&  1/24 \\
\bar J_3 &=& 1/108 \\
\bar J_4 &=&-\; 5/576 \\
\bar J_5 &=&  - \; 11/2160 \\
\bar J_6 &=&  175/46656
\end{eqnarray}

The results above for $\bar J_i, \ i \le 4$, were presented in [1] as obtained by hand computation.

The  $\bar J_i$ for $d=1$ are computable in seconds of computer time, and  $\bar J_i$ with much higher values of $i$ are easily obtained.  For $d=2$ and $d=3$ the  $\bar J_6$ required several days of computer time;  $\bar J_7$ would be extremely difficult to obtain, requiring both more theory and much more computer time, a year's effort.

The results, (4) through (21), and equation (30) of [1] (whose proof will appear in a later paper), will establish the following formulae for  $\bar J_i$ in general dimension:
\begin{eqnarray}
\bar J_1 &=& 0 \\
\bar J_2 &=&  \frac 18 \; \frac 1 d \\
\bar J_3 &=& \frac 1{12}\; \frac1{d^2} \\
\bar J_4 &=&-\; \frac 3{32} \; \frac1{d^2} + \frac3{64} \; \frac 1{d^3} \\
\bar J_5 &=&  - \; \frac 1 8 \; \frac 1{d^3} - \frac 3{80} \; \frac 1 {d^4} \\
\bar J_6 &=&  \frac 7{48}  \frac1{d^3} -  \frac 5{64} \frac 1{d^4} -  \frac 1 6 \frac1{d^5}
\end{eqnarray}

Substituting these values of $\bar J_i$ into (36) of [1], and extracting the behavior of the exponent in (36) to order $1/d^3$, we get the result in (2) above.  We are using the fact that $\bar J_i$ for $i > 6$ have only powers of $1/d$ greater than 3.

For $d=2$ and $d=3$ we have considered two sequences of approximations each characterized by keeping more and more terms in some sense.  The first sequence of approximations
\be	A_0, A_1, \cdots  \ee
is constructed from
\be    \la_d \sim \ \frac 1 2 \ ln(2d) - \frac 1 2 + \frac{c_1}d + \frac{c_2}{d^2} + \cdots  \ee
by
\be	A_r = \frac 1 2 \ ln(2d) - \frac 1 2 + \sum^r_{i=1} \ \frac{c_i}{d^i} . \ee
We then have the results:

\underline{For $d=2$}:
\begin{eqnarray}
A_0 &=& .1931 \\
A_1 &=& .2556 \\
A_2 &=& .2687 \\
A_3 &=& .2784 
\end{eqnarray}
Of course here, for $d=2$, the exact result $\la_2$ = .29156.... is known. [3], [4]

\underline{For $d=3$}:
\begin{eqnarray}
A_0 &=& .3959 \\
A_1 &=& .4375 \\
A_2 &=& .4433 \\
A_3 &=& .4462 
\end{eqnarray}
Exact bounds for $\la_3$ as given in [5] are 
\be   .440075 \le \la_3 \le .457547  \ee

Our second sequence of approximations is
\be	B_0, B_1, \cdots	\ee
constructed by replacing $\bar J_i$ by $\bar J_i \; x^{i-1}$ with $x$ a formal parameter.  We then substitute these  
$\bar J_i \; x^{i-1}$ for $\bar J_i $ in (36) of [1].  To obtain $B_r$, keeping terms through power $x^r$ (and then setting $x=1$), we get the values:

\underline{for d = 2}:
\begin{eqnarray}
B_0 &=& .1931  \\
B_1 &=& .2556  \\
B_2 &=& .2921  \\
B_3 &=& .2992 \\
B_4 &=& .2905  \\
B_5 &=& .2814
\end{eqnarray}

\underline{and for d = 3}:
\begin{eqnarray}
B_0 &=& .3959  \\
B_1 &=& .4375  \\
B_2 &=& .4538  \\
B_3 &=& .4524  \\
B_4 &=& .4468  \\
B_5 &=& .4445
\end{eqnarray}

Originally we hoped the sequence of $B_i$ converged, but this now seems unlikely.  If the sequences of $B_i$ for $d=2$ and $d=3$ behaved similarly, we might have an argument that $\la_3 = .453 \pm .001$.  We do not know whether to believe this or not.

It remains to outline the general flow of the computer computations.  The quantities of interest were obtained through a chain of operations successively performed.  Were we to redo the hand calculation of $\bar J_4$ as described in [1], we would find it much more efficient to echo some of the ideas used in the computer programs.  Naturally we cannot describe the many little clever ideas, saving memory and increasing efficiency, needed to push the computations to the level of obtaining $\bar J_6$.  We find it convenient to describe our computations as divided into four steps, which we present in inverse order.

\noindent
\underline{Step 4}.  This step assumes knowledge of the $\bar J_i$, and substitutes them into (36) of [1] to obtain the approximations $A_i$ and $B_i$.

\noindent
\underline{Step 3}.  This step assumes known the quantities
\be	 \sum_{s_1,s_2,...,s_s} \ f(\bar s_1) \cdots f(\bar s_s) \psi'_c(s_1,s_2,...,s_s)	\ee
(where $f$ is as in [1], $\frac 1{2d}$ on a dimer, zero on other tiles, and (53) is as (28) of [1] with $v$ replaced by $f$) and computes
\be	J_s = \sum_{s_1,s_2,...,s_s} \ v(\bar s_1) \cdots v(\bar s_s) \psi'_c(s_1,s_2,...,s_s).	\ee

\noindent
\underline{Step 2}.  Takes the output of Step 1 described next to yield the expression in equation (53) above.

Note that in equations (53) and (54) above the sums are over all located tiles $s_1,...,s_s$ such that the sequence is connected.  That is, one cannot pick two subsets of ordered sets $s_1,...,s_s$ such that, with the two sub-ordered sequences
\[	s_{\al_1}\; , ..., \; s_{\al_r} \ \ {\rm and} \ \ s_{\beta_1}\; , ..., \; s_{\beta_r}	\]
one has 
\begin{itemize}
\item [1)] the same $s_i$ appear in $s_1,...,s_s$ and in the union of the two sub-ordered sets, with the same multiplicity
\item [2)] the two sub-ordered sets of located tiles have no overlap between them.
\end{itemize}
We now define an $s$-{\it dimer tree} to be an ordered sequence of $s$ located tiles $s_1, s_2, ..., s_s$ such that for each $i \le s$ the sequence of located tiles $s_1, s_2, ..., s_i$ is connected.  It is easy to see $s$-{\it dimer trees} are the sort of thing a computer can easily generate.

\noindent
\underline{Step 1}.  The output of Step 1 (towards computation of $J_s$) is the set of all possible $s$-{\it dimer trees} with given $s_1$.  With each $s$-{\it dimer tree} the output of Step 1 also includes knowledge of the overlap pattern of the tree.  That is, for the tree $s_1,...,s_s$ one knows whether $s_i \cap s_j$ is empty or not for each $i$ and $j$.  (The output of Step 1 actually yields the number of $s$-{\it dimer trees} with each overlap pattern.)  

Steps 3 and 4 take seconds to run in the cases we have studied.  In the computation of $\bar J_6$, Step 1 and Step 2 required several weeks of computer time.

\bigskip
\bigskip
\bigskip
\centerline{References}
\begin{itemize}
\item[[1]] Paul Federbush, Hidden Structure in Tilings, Conjectured Asymptotic Expansion for $\la_d$ in Multidimensional Dimer Problem, \ arXiv : 0711.1092V9 [math-ph].
\item[[2]] Henryk Minc, An Asymptotic Solution of the Multidimensional Dimer Problem, Linear and Multilinear Algebra, 1980, {\bf 8}, 235-239.
\item[[3]] E.M. Fisher, Statistical Mechanics of Dimers on a Plane Lattice, Phys. Rev. 124 (1961), 1664-1672.
\item[[4]] P.W. Kasteleyn, The Statistics of Dimers on a Lattice, Physica 27 (1961), 1209-1225.
\item[[5]] S. Friedland, E. Krop, P.H. Lundow, K. Markstr\"{o}m, Validations of the Asymptotic Matching Conjectures, Math/0603001v2.

\end{itemize}

\end{document}